**POSTURAL DESTABILIZATION INDUCED BY TRUNK EXTENSOR MUSCLES FATIGUE IS**

**SUPPRESSED BY USE OF A PLANTAR PRESSURE-BASED ELECTRO-TACTILE BIOFEEDBACK**

Nicolas VUILLERME[1], Nicolas PINSAULT[1], Olivier CHENU[1], Anthony FLEURY[1], Yohan PAYAN[1] and Jacques DEMONGEOT[1]

[1] Laboratoire TIMC-IMAG, UMR UJF CNRS 5525, La Tronche, France

**Address for correspondence:**

Nicolas VUILLERME

Laboratoire TIMC-IMAG, UMR UJF CNRS 5525

Faculté de Médecine

38706 La Tronche cédex

France.

Tel: (33) (0) 4 76 63 74 86

Fax: (33) (0) 4 76 51 86 67

Email: nicolas.vuillerme@imag.fr

**Abstract**

Separate studies have reported that postural control during quiet standing could be (1) impaired with muscle fatigue localized at the lower back, and (2) improved through the use of plantar pressure-based electrotactile biofeedback, under normal neuromuscular state. The aim of this experiment was to investigate whether this biofeedback could reduce postural destabilization induced by trunk extensor muscles. Ten healthy adults were asked to stand as immobile as possible in four experimental conditions: (1) no fatigue/no biofeedback, (2) no fatigue/biofeedback, (3) fatigue/no biofeedback and (4) fatigue/biofeedback. Muscular fatigue was achieved by performing trunk repetitive extensions until maximal exhaustion. The underlying principle of the biofeedback consisted of providing supplementary information related to foot sole pressure distribution through electro-tactile stimulation of the tongue. Centre of foot pressure (CoP) displacements were recorded using a force platform. Results showed (1) increased CoP displacements along the antero-posterior axis in the fatigue than no fatigue condition in the absence of biofeedback and (2) no significant difference between the no fatigue and fatigue conditions in the presence of biofeedback. This suggests that subjects were able to efficiently integrate an artificial plantar pressure information delivered through electro-tactile stimulation of the tongue that allowed them to suppress the destabilizing effect induced by trunk extensor muscles fatigue

## INTRODUCTION

A biofeedback system whose underlying principle consists of providing supplementary information related to foot sole pressure distribution through electro-tactile stimulation of the tongue has recently been developed for balance control (Vuillerme et al. 2007b, c, d). Before testing with balanceimpaired patients, it was first important to determine whether this biofeedback system could improve postural control in individuals with intact sensory, motor, cognitive capacities. Within this context, previous experiments conducted in young healthy adults have reported reduced centre of foot pressure (CoP) displacements when this biofeedback was in use relative to when it was not, hence evidencing the ability of the central nervous system (CNS) to efficiently integrate an artificial plantar-based, tongue-placed electro-tactile biofeedback for controlling posture during quiet standing (Vuillerme et al. 2007b, c, d). The above-mentioned investigations have been performed under a normal neuromuscular state, in conditions which did not endanger postural stability. At this point, muscle fatigue, which represents an inevitable phenomenon for physical and daily activities, is one factor that could affect the integrity of the neuromuscular system. Indeed, it is recognized that muscle fatigue impairs the peripheral proprioceptive system, the central processing of proprioception and the force-generating capacity (e.g., Taylor et al. 2000). Accordingly, postural control being considered as a sensor-motor process (e.g., Schmidt 1975), a decreased postural control during bipedal quiet standing has been reported following lower limbs efforts (e.g. Ledin et al. 2004; Vuillerme et al. 2002; Vuillerme et al. 2006a; Vuillerme and Demetz 2007). Recent studies also have documented decreased postural control following localized muscle fatigue at the lower back (Davidson et al. 2004; Madigan et al. 2006; Pline et al. 2006; Vuillerme et al. 2007a; Vuillerme and Pinsault 2007), stressing the importance of intact lumbar muscle function on the control of posture during quiet standing.

The present study thus aimed at assessing the effects of a plantar pressure-based electro-tactile biofeedback following trunk extensor muscles fatigue. Precisely, we investigated whether such a biofeedback could reduce the postural destabilization induced by trunk extensor muscles fatigue, i.e., to compensate from an alteration of the neuromuscular function at the lower back. It was hypothesised that: (1) without the provision of the biofeedback, trunk extensor muscles fatigue would increase CoP displacements (Davidson et al. 2004; Madigan et al. 2006; Pline et al. 2006; Vuillerme et al. 2007a; Vuillerme and Pinsault 2007), and (2) the availability of the biofeedback would reduce the destabilizing effect induced by trunk extensor muscles fatigue.

**METHODS**

Ten healthy male university students (age: $25.2 \pm 3.2$ years; body weight: $77.2 \pm 5.2$ kg; height: $181.0 \pm 3.6$ cm; mean $\pm$ SD) with no motor problems, neck injury, vertigo, neurological disease, or vestibular impairment voluntarily participated in the experiment. They gave their informed consent to the experimental procedure as required by the 1964 Helsinki declaration and the local Ethics Committee.

Eyes closed, subjects stood barefoot in a natural position (feet abducted at 30°, heels separated by 3 cm), their arms hanging loosely by their sides and were asked to sway as little as possible.

This postural task was executed under two experimental conditions of No Biofeedback and Biofeedback. The No Biofeedback condition served as a control condition. In the Biofeedback condition, subjects performed the postural task using a plantar pressure-based, tongue-placed electro-tactile biofeedback system (Vuillerme et al. 2007b,c,d). A plantar pressure data acquisition system (FSA Inshoe Foot pressure mapping system, Vista Medical Ltd, Winnipeg, Manitoba, Canada), consisting of a pair of 2 mm thick flexible insoles instrumented with an array of $8 \times 16$ pressure sensors per insole (1cm² per sensor, range of measurement: 0-30 PSI), was used. The pressure sensors transduced the magnitude of pressure exerted on each left and right foot sole at each sensor location into the calculation of the positions of the resultant ground reaction force exerted on each left and right foot, referred to as the left and right foot centre of foot pressure (CoP), respectively ($CoP_{lf}$ and $CoP_{rf}$). The positions of the resultant CoP were then computed from the left and right foot CoP trajectories through the following relation (Winter et al. 1996):

$$CoP = CoP_{lf} \times R_{lf} / (R_{lf} + R_{rf}) + CoP_{rf} \times R_{rf} / (R_{rf} + R_{lf}),$$

where $R_{lf}$, $R_{rf}$, $CoP_{lf}$, $CoP_{rf}$ are the vertical reaction forces under the left and the right feet, the positions of the CoP of the left and the right feet, respectively.

CoP data were then fed back in real time to a recently developed tongue-placed tactile output device (Vuillerme et al. 2006b, 2007a, b, c, d). This so-called Tongue Display Unit (TDU), initially introduced by Bach-y-Rita et al. (1998; Bach-y-Rita and Kercel 2003) comprises a 2D array (1.5 9 1.5 cm) of 36 electro-tactile electrodes each with a 1.4 mm diameter, arranged in a 6 9 6 matrix. The matrix of electrodes, maintained in close and permanent contact with the front part of the tongue dorsum, was connected to an external electronic device triggering the electrical signals that stimulate the tactile receptors of the tongue via a flat flexible cable passing out of the mouth (Fig. 1). Note that unipolar electrodes were used; only one electrode was activated at a time and the unpulsed ones served as the return current path.

-------------------------------------

Please insert Figure 1 about here

-------------------------------------

The underlying principle of the biofeedback system was to supply subjects with supplementary information about the position of the CoP relative to a predetermined adjustable ''dead zone'' (DZ) through the TDU. In the present experiment, antero-posterior and medio-lateral bounds of the DZ were set as the standard deviation of subject's CoP displacements recorded for 10 s preceding each experimental trial. To avoid an overload of sensory information presented to the user, a simple and intuitive coding scheme for the TDU, consisting in a ''thresholdalarm'' type of feedback rather than a continuous feedback about ongoing position of the CoP, was used. As illustrated in Fig. 2, (1) when the position of the CoP (white triangle) was determined to be within the DZ (grey rectangle), no electrical stimulation was provided in any of the electrodes of the matrix; (2) when the position of the CoP was determined to be outside the DZ - i.e., when it was most needed - electrical stimulation was provided in distinct zones of the matrix (black dots), depending on the position of the CoP relative to the DZ. Specifically, four different zones located in the front, rear, left and right portion of the matrix were defined; the activated zone of the matrix corresponded to the position of the CoP relative to the DZ. For instance, in the case that the CoP was located at the right-hand side of the DZ, a stimulation of four electrodes located in the right portion of the matrix (i.e. stimulation of the right portion of the tongue) was provided.

-------------------------------------

Please insert Figure 2 about here

-------------------------------------

The intensity of the electrical stimulating current was adjusted for each subject, and for each of the front, rear, left, and right portions of the tongue. Several practice runs were performed prior to the test to ensure that subjects had mastered the relationship between the position of the CoP relative to the DZ and lingual stimulations. Note that the foot insole system was put beneath the feet and the TDU was inserted in the oral cavity of the subject during all trials of the experiment (i.e., in both the no biofeedback and biofeedback conditions), ruling out the possibility the postural improvement observed in the biofeedback relative to the no biofeedback condition to be due to enhanced plantar cutaneous facilitation and mechanical stabilization of the head in space, respectively.

All trials were performed during a single experimental session. The two no biofeedback and biofeedback conditions were executed before (no fatigue condition) and immediately after a designated fatiguing exercise for trunk extensor muscles (fatigue condition). The order of presentation of the two no biofeedback and biofeedback conditions was randomised over subjects. The muscular fatigue was induced until maximal exhaustion with trunk repetitive extensions (Vuillerme et al. 2007a; Vuillerme and Pinsault 2007). As illustrated in Fig. 3, subjects lay prone on an horizontal bench with their upper body unsupported parallel to the ground, their lower extremities secured to the bench with straps at the hips, knees and ankles and their arms hold crossed the chest. Subjects were instructed to raise their upper body to a horizontal position and then lowering it back down (i.e., trunk extension exercise through a 90° range of motion) as many times as possible following the beat of a metronome (40 beats/min).

------------------------------------

Please insert Figure 3 about here

------------------------------------

The examiner gave verbal encouragement before and during each contraction. The fatigue level was reached when subjects were no more able to complete the trunk extension exercise. Immediately on the cessation of exercise, the subjective exertion level was assessed through the Borg CR-10 scale (Borg 1990). Mean Borg ratings of $8.5 \pm 0.7$ and $8.7 \pm 0.7$ were recorded for the fatigue/no biofeedback and fatigue/biofeedback conditions, respectively. This indicates that subjects rated their perceived fatigue in the trunk extensor muscles between ''very strong'' (Borg rating of seven) (Borg 1990) and ''extremely strong'' (Borg rating of ten) (Borg 1990). The recovery process after fatigue procedures is often considered as a limitation for all fatigue experiments. In the present study, to ensure that balance measurement in the fatigue condition was obtained in a genuine fatigued state, various rules were respected (Vuillerme et al. 2007a; Vuillerme and Pinsault 2007). (1) The fatiguing exercise took place beside the force platform, so that there was a short time-lag between the exercise-induced fatiguing activity and the balance measurements (less than 1 min) and (2) the fatiguing exercise was repeated prior to each trial. For each condition of no biofeedback and biofeedback and each condition of no-fatigue and fatigue, subjects performed three 30-s trials, for a total of 12 trials.

A force platform (Equi+, model PF01, Aix les Bains, France), which was not a component of the biofeedback system, was used to measure the displacements of the CoP (64 Hz sampling frequency) as a gold-standard system for assessment of postural control during upright quiet standing.

CoP displacements were processed through a space-time domain analysis including the calculation of the surface area covered by the trajectory of the CoP with a 90% confidence interval (Tagaki et al. 1985) and the variances of positions of the CoP along the medio-lateral (ML) and antero-posterior (AP) axes.

The means of the three trials performed in each of experimental condition were used for statistical analyses.

Data obtained for the surface area covered by the trajectory of the CoP were submitted to a 2 Fatigues (No Fatigue *vs.* Fatigue) × 2 Biofeedback (No Biofeedback *vs.* Biofeedback) analysis of variance (ANOVA) with repeated measures on both factors. To further investigate whether the effects of Fatigue and Biofeedback were similar according to the ML or AP axes, a 2 Fatigues (No Fatigue *vs.* Fatigue) × 2 Biofeedback (No Biofeedback *vs.* Biofeedback) × 2 Axes (Medio-lateral *vs.* Antero-posterior) ANOVA with repeated measures on all factors was applied to the variance of the CoP displacements. Post-hoc analyses (*Newman-Keuls*) were performed whenever necessary. Level of significance was set at 0.05.

**RESULTS**

Figure 4 illustrates representative CoP displacements from a typical subject recorded in the four experimental conditions of 1) No Fatigue / No Biofeedback, (2) No Fatigue / Biofeedback, (3) Fatigue / No Biofeedback and (4) Fatigue / Biofeedback.

------------------------------------

Please insert Figure 4 about here

------------------------------------

Surface area covered by the trajectory of the CoP

Analysis of the surface area covered by the trajectory of the CoP showed a significant interaction of Fatigue × Biofeedback ($F(1,9) = 6.99$, $P < 0.05$). As illustrated in Figure 5, the decomposition of this interaction into its simple main effects indicated that (1) fatigue increased CoP surface area in the absence of biofeedback ($P < 0.001$), and (2) CoP surface area was not affected by fatigue in the presence of biofeedback ($P > 0.05$). The ANOVA also showed a significant main effects of Fatigue ($F(1,9) = 8.75$, $P < 0.05$) and Biofeedback ($F(1,9) = 6.18$, $P < 0.05$).

------------------------------------

Please insert Figure 5 about here

------------------------------------

Variance of the CoP displacements along the ML and AP axes

Analysis of the variance of the CoP displacements showed a significant three-way interaction of Fatigue × Biofeedback × Axis ($F(1,9) = 5.27$, $P < 0.05$). As illustrated in Figure 6, the decomposition of this interaction into its simple main effects indicated that (1) fatigue increased variance of the CoP displacements along the AP axis in the absence of biofeedback ($P < 0.001$), and (2) the variances the CoP displacements along both the ML and AP axes were not affected by fatigue in the presence of biofeedback ($Ps > 0.05$). The ANOVA also showed a significant main effects of Fatigue ($F(1,9) = 8.95$, $P < 0.05$), Biofeedback ($F(1,9) = 6.08$, $P < 0.05$) and Axis ($F(1,9) = 15.19$, $P < 0.01$).

------------------------------------

Please insert Figure 6 about here

------------------------------------

DISCUSSION

The present study was designed to investigate whether a plantar pressure-based electro-tactile biofeedback (Vuillerme et al. 2007b, c, d) could reduce the postural destabilization induced by trunk extensor muscles fatigue.

Without the provision of biofeedback (no biofeedback condition), results showed a wider surface area covered by the trajectory of the CoP in the fatigue than no fatigue condition (Fig. 5). This result supports our hypothesis 1, in accordance with recent observations (Davidson et al. 2004; Madigan et al. 2006; Pline et al. 2006; Vuillerme et al. 2007a; Vuillerme and Pinsault 2007). Analysis of the variance of the CoP displacements further indicated that the destabilizing effect of Fatigue was more accentuated along the AP than ML axis (Fig. 6). On the whole, these results confirm the important role of lumbar neuromuscular system in postural control during quiet standing. Conversely, when the biofeedback was available (Biofeedback condition), CoP displacements were not affected by trunk extensor muscles fatigue, as indicated by the significant interactions of fatigue × biofeedback and fatigue × biofeedback × axis, for the surface area covered by the trajectory of the CoP (Fig. 5) and the variance of the CoP displacements (Fig. 6), respectively. In other words, supporting our hypothesis 2, the availability of the biofeedback allowed the subjects not only to reduce, but also to suppress the destabilizing effect induced by trunk extensor muscles fatigue.

At this point, a possible reason leading to these results could be that the use of the tongue-placed electro-tactile biofeedback may have lead subjects to pay more attention to the regulation of their postural oscillations. However, in a recent study, in which subjects were instructed to deliberately focus their attention on their body sway and to increase their active intervention into postural control, postural oscillations were not reduced (Vuillerme and Nafati 2007). We thus believe that the postural improvement observed in the Biofeedback condition could not be attributed to the subjects' paying more attention to the regulation of their postural sway. Our results rather evidence the ability of the CNS to efficiently integrate an artificial plantar pressure information delivered through electro-tactile stimulation of the tongue to improve postural control (Vuillerme et al. 2007b, c, d), and to compensate for a postural disturbance induced by an alteration of neuromuscular function at the lower back. Note that such adaptive process ensuring a stabilization of individuals' postural behaviour in conditions of muscles fatigue previously has been reported (e.g., Ledin et al. 2004; Vuillerme et al. 2005; Vuillerme et al. 2006a; Vuillerme and Demetz 2007). Interestingly, with regard to the hypothesis of an increase of the quality of somatosensory from the plantar soles induced by the use the plantar-based electrotactile biofeedback, results of the present experiment are in line with those of a recent study reporting that a decreased destabilizing effect of trunk extensor muscles fatigue observed under normal somatosensation from the foot was facilitated by providing increased cutaneous feedback at the foot and ankle (Vuillerme and Pinsault 2007). From now on, to assess the potential clinical value of the plantar pressurebased electro-tactile biofeedback system in enhancing/restoring/preserving balance in balance-impaired subjects, further testing of individuals with reduced neuromuscular function and/or sensorimotor capacities - resulting either from normal aging, trauma or disease - is warranted.

**ACKNOWLEDGEMENTS**

We are indebted to Professor Paul Bach-y-Rita for introducing us to the Tongue Display Unit and for discussions about sensory substitution. Paul has been for us more than a partner or a supervisor: he was a master inspiring numerous new fields of research in many domains of neurosciences, biomedical engineering and physical rehabilitation. The authors would like to thank subject volunteers and the anonymous reviewers for their valuable comments and suggestions on the first version of the manuscript. The company Vista Medical is acknowledged for supplying the FSA Inshoe Foot pressure mapping system. This research was supported by the company IDS and the Fondation Garches.

**Figure captions**

**Figure 1.** Photograph of the Tongue Display Unit used in the present experiment. It comprises a 2D electrode array (1.5 × 1.5 cm) consisting of 36 gold-plated contacts each with a 1.4 mm diameter, arranged in a 6 × 6 matrix.

**Figure 2.** Sensory coding schemes for the Tongue Display Unit (TDU) as a function of the position of the centre of foot pressure (CoP) relative to a predetermined dead zone (DZ). White triangles, grey rectangles and black dots represent the positions of the CoP, the predetermined dead zones and activated electrodes, respectively. On the one hand, no electrodes were activated when the CoP position was determined to be within the DZ (central panel). One the other hand, four electrodes located in the front, rear, left, and right zones of the matrix of the TDU were activated when the CoP positions were determined to be outside the DZ, located towards the front, rear, left and right of the DZ, respectively (peripheral panels). These four stimulation patterns correspond to the stimulations of the front, rear, left and right portions of the tongue dorsum, respectively.

**Figure 3.** Experimental setup. Subjects lay prone on an horizontal bench with their upper body unsupported parallel to the ground, their lower extremities secured to the bench with straps at the hips, knees and ankles and their arms hold crossed the chest. Subjects were asked to perform trunk repetitive extensions until exhaustion. This exercise was performed through a 90° range of motion, with full extension being parallel to the ground. A metronome set at 40 beats/min was used to ensure appropriate and consistent timing.

**Figure 4.** Representative displacements of the centre of foot pressure (CoP) from a typical subject recorded in the four (1) No Fatigue / No Biofeedback, (2) No Fatigue / Biofeedback, (3) Fatigue / No Biofeedback and (4) Fatigue / Biofeedback conditions.

**Figure 5.** Mean and standard error of surface area covered by the trajectory of the CoP obtained for the two conditions of No Fatigue and Fatigue of trunk extensor muscles and the two conditions of No Biofeedback and Biofeedback. The two conditions of No Biofeedback and Biofeedback are presented with different symbols: No Biofeedback (*white bars*) and Biofeedback (*black bars*). The significant $P$ values for comparisons between the No Biofeedback and Biofeedback conditions are also reported (*: $P < 0.05$; ***: $P < 0.001$).

**Figure 6.** Mean and standard error of the variance of the CoP displacements along the medio-lateral and antero-posterior axes obtained for the two conditions of No Fatigue and Fatigue of trunk extensor muscles and the two conditions of No Biofeedback and Biofeedback. The two conditions of No Biofeedback and Biofeedback are presented with different symbols: No Biofeedback (*white bars*) and Biofeedback (*black bars*). The significant $P$ values for comparisons between the No Biofeedback and Biofeedback conditions are also reported (*: $P < 0.05$; ***: $P < 0.001$).

**Figure 1**

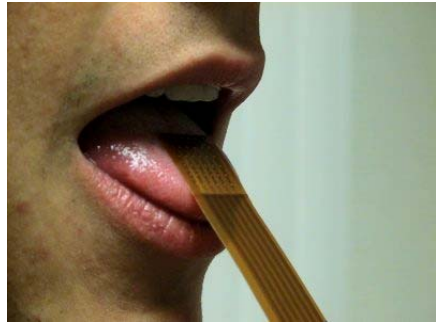

**Figure 2**

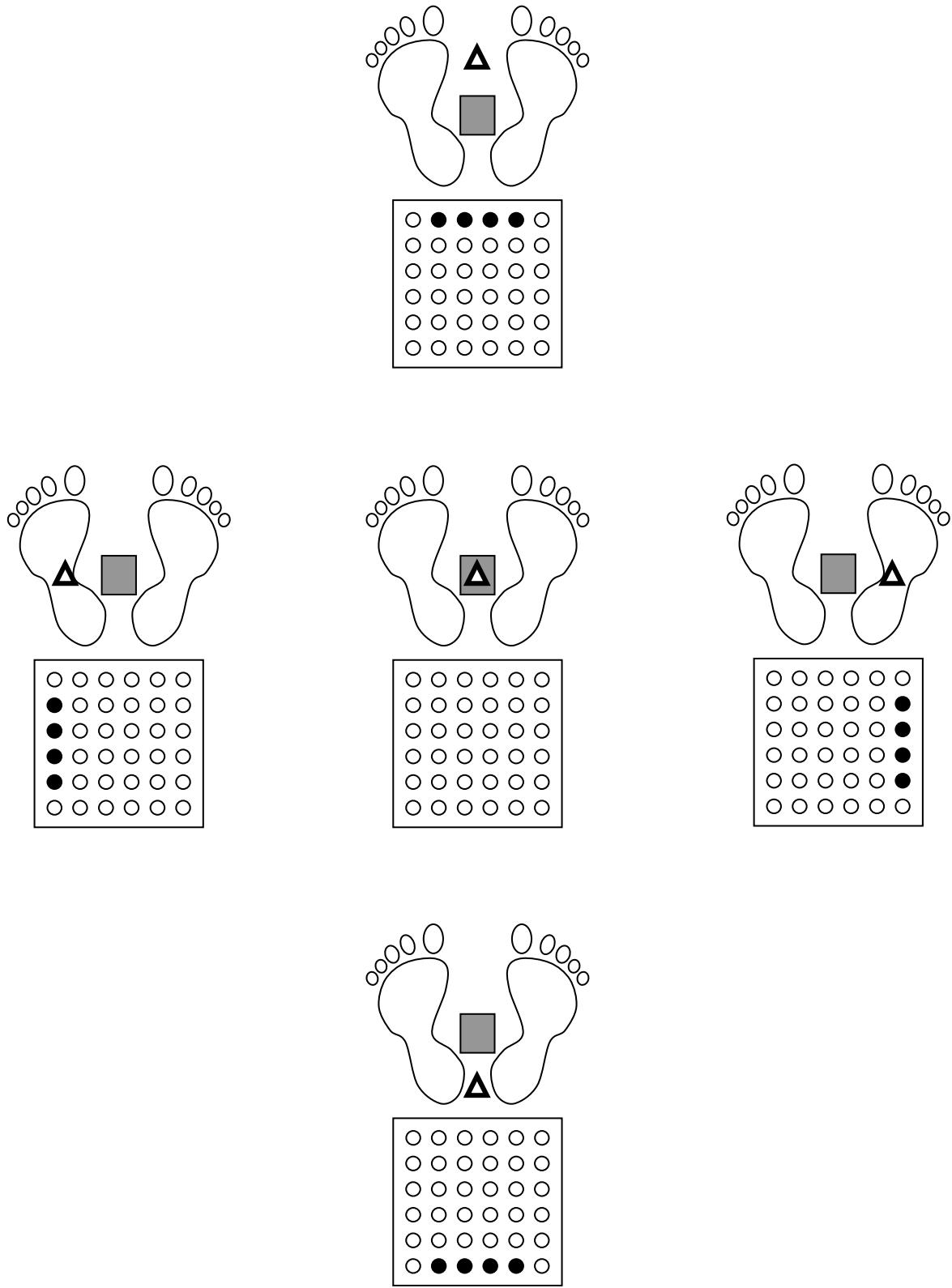

**Figure 3**

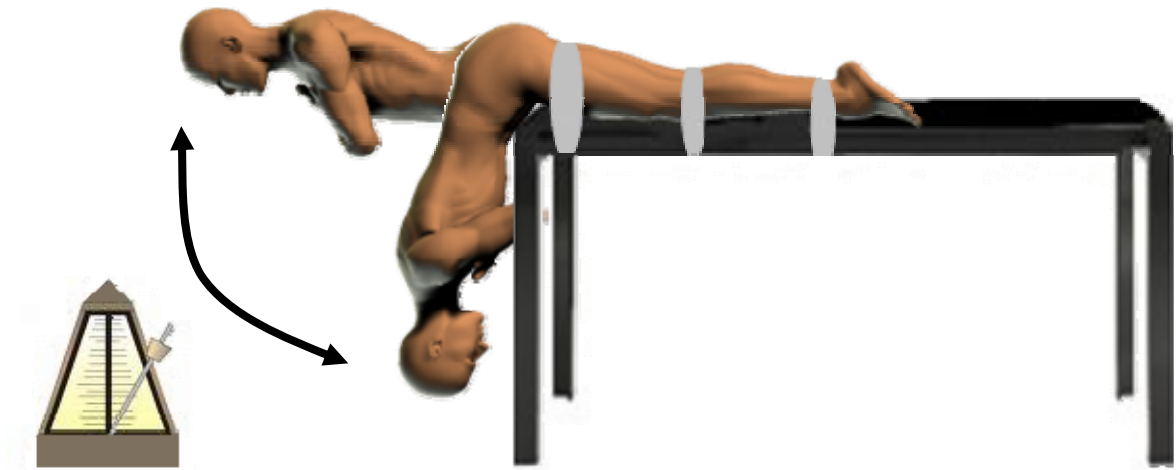

**Figure 4**

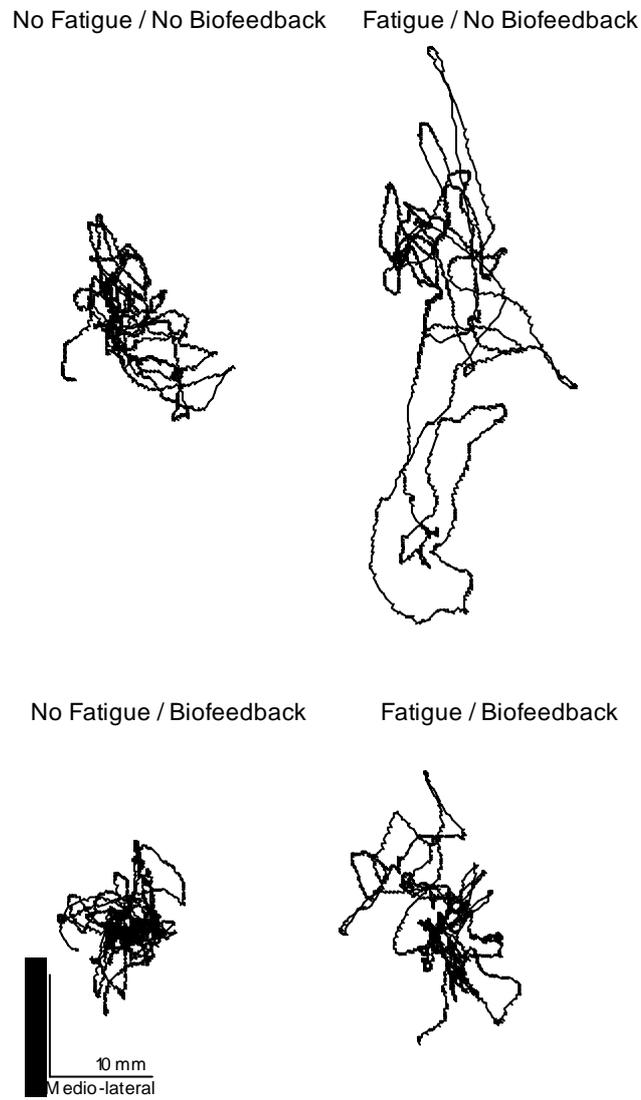

**Figure 5**

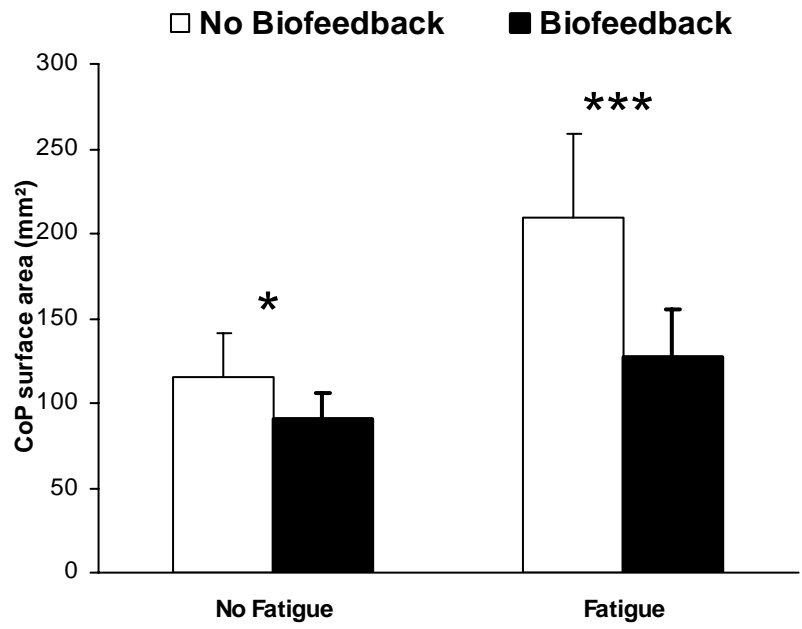

**Figure 6**

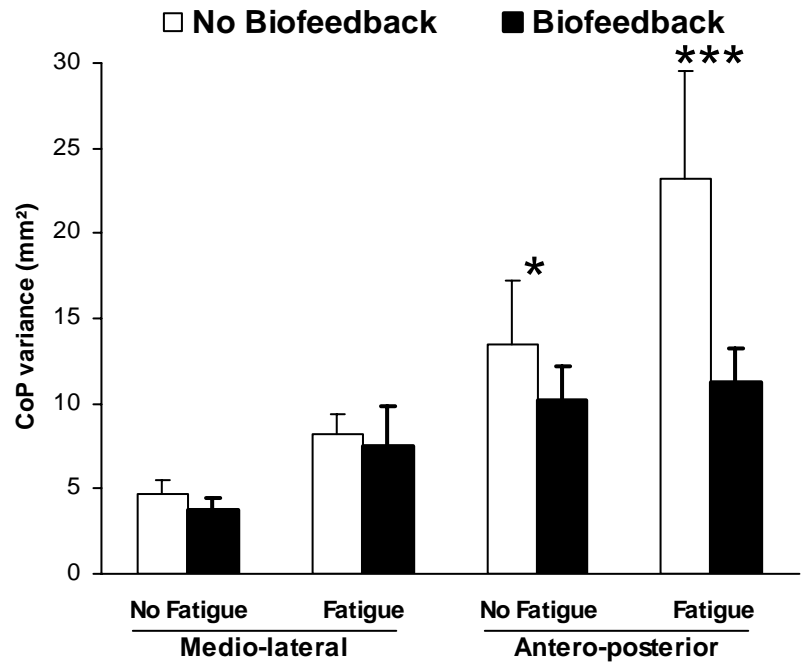